\documentclass[twocolumn]{revtex4}
\usepackage{graphics}
\usepackage{amssymb,epsf}
\usepackage{amsmath,mathrsfs}
\usepackage{graphicx,epsfig}
\usepackage{latexsym,amssymb}
\usepackage{epsf,longtable,array}
\usepackage{ifpdf,natbib,lineno}

\begin{document}
\title{\Large{\textbf{Dynamics of global entanglement under decoherence }}}
\author{Afshin Montakhab\footnote[1]{E-mail address: montakhab@shirazu.ac.ir}
and Ali Asadian}
\address{Physics Department, College of Sciences, Shiraz University, Shiraz 71454, Iran.}
\begin{abstract}
We investigate the dynamics of global entanglement, the Meyer-Wallach
measure, under decoherence, analytically. We study two important class of
multi-partite entangled states, the Greenberger-Horne-Zeilinger and the W state.
We obtain exact results for various models of system-environment interactions
(decoherence). Our results shows distinctly different scaling behavior for these
initially entangled states indicating a relative robustness of the W state, consistent
with previous studies. \\
\\ PACS number(s): 03.67.Mn, 03.65.Ud, 03.65.Yz
\end{abstract}
\maketitle

\section{Introduction}

It is well-known that the notion of
quantum entanglement is a key concept in quantum mechanics. It is
also responsible for \lq\lq strange" non-local behavior of quantum
systems in marked contrast to classical notions of reality and
locality [1,2]. Schr\"{o}dinger refer to it as the \lq\lq essence
of quantum mechanics" [1]. Besides its fundamental aspects,
entanglement constitutes the central part of new modes of
information technology, quantum computation and quantum
communication [3,4,5] and is therefore a key ingredient of many
information processing protocols. Recently, a considerable amount
of work has been devoted to characterize, quantify and realize
different variety of entangled states [6]. By now, bi-partite
entanglement is a relatively well-understood phenomena. However,
the situation becomes much more complex when multi-partite systems
are considered [7].

On the other hand, entangled states are very fragile when they are
exposed to environment. The biggest enemy of entanglement is
decoherence which is believed to be the responsible mechanism for
emergence of the classical behavior in quantum systems [8]. Since
the maintenance and control of entangled states is essential to
realization of quantum information processing systems, the study of
deteriorating effect of decoherence in entangled states would be of
considerable importance from theoretical as well as experimental
point of view [9,10,11].

To demonstrate the effects of decoherence on entangled states, an
appropriate entanglement measure which could be capable of
monitoring the dynamics of entanglement in decoherence processes is
needed . However, there are no exact measure of entanglement under
general conditions for mixed states. Even for bipartite mixed
states, apart from the particular case of two-level systems [12],
the exact solution is missing. There is an approximate
generalization of concurrence for mixed states which was proposed by
Mintert et al [13] and has bean used to measure entanglement in
multi-qubit systems.

Global entanglement(GE), defined by Meyer-Wallach (MW) entanglement
measure of pure-state [14] which is a monotone[15], is a very useful
measure of entanglement. As we will show briefly, GE is a measure of
total non-local information per particle in a general multi-qubit
system. Therefore, GE gives an intuitive meaning to multi-qubit
entanglement as well being an experimentally accessible measure
[16]. In this paper, we use this measure to monitor the entanglement
dynamics of two seminal multi-qubit entangled states, the
Greenberger-Horne-Zeilinger (GHZ) state and the W state, under
different models of system-environment interaction.

To demonstrate the GE dynamics of multi-qubit entangled state
under decoherence, we need to know the generalization of GE to
mixed state. Unfortunately, there is no generalization of the primary
definition of the MW measure to mixed states, analytically. However, we
can monitor the GE dynamics for two important classes of
multi-qubit entangled state, the GHZ and W state, by exploiting
the relationship between GE and tangles. To elucidate this point,
we adopt informational approach which can give an intuitive
meaning to GE.

\section{Global Entanglement, Information, and Tangles}

Finite amount of information can be attributed to N-qubit pure state
which is N bit of information according to Brukner-Zeilinger
operationally invariant information measure [17]. This information
can be distributed in local as well as non-local form, which is
associated with entanglement [18].This information has a
complimentary relation:
\begin{equation}
 I_{total}=I_{local}+I_{non-local}.
\end{equation}
The total information is conserved unless transferred to environment
through decoherence. The amount of information in local form is
$I_{local}=\sum_{i=1}^{N}I_{i}$ where, $I_{i}=2Tr\rho_{i}^{2}-1$ is
the operationally invariant information measure of a qubit [17].
Therefore, according to Eq.(1)
$I_{non-local}=\sum_{i=1}^{N}2(1-Tr\rho_{i}^{2})$ which can be
distributed in different forms of quantum correlations, the tangles,
among the system,
\begin{equation}
 I_{non-local}=2\sum_{i_{1}<i_{2}}\tau_{i_{1}i_{2}}+...+N\sum_{i_{1}<i_{2}<...<i_{N}}\tau_{i_{1}...i_{N}},
\end{equation}
where the first term is referred to as 2-tangle, the next being
3-tangle and the last term the N-tangle of the system. It can
easily be seen that the MW measure of GE can be written as:
\begin{eqnarray}
 E_{gl}=\frac{1}{N}[2\sum_{i_{1}<i_{2}}\tau_{i_{1}i_{2}}+...+N\sum_{i_{1}<i_{2}<...<i_{N}}\tau_{i_{1}...i_{N}}].
\end{eqnarray}
Therefore, the MW measure of GE ($E_{gl}$) is a measure of total
non-local information per particle (or average of tangles per
particles $\frac{\langle\tau\rangle}{N}$). This property of $E_{gl}$
resembles the molar-extensive thermodynamic variables (e.g. heat
capacity $C_{p}$). In fact, we expect $E{gl}$ to have
thermodynamically relevant and experimentally accessible features in
multi-qubit (spin) systems[19]. This property of GE in contrast to
most other multi-partite entanglement measures. It is therefore of
considerable interest to investigate the dynamics of GE under
decoherence in the large N limit for the generic, experimentally
realizable, entangled states.  We note that $E_{gl}$ does not give
detailed knowledge of tangles distribution among the system. For
example, $E_{gl}$ cannot distinguish between entangled states  which
have equal $\langle\tau\rangle$ yet different distribution of
tangles, like $|GHZ\rangle_{N}$ and $|EPR\rangle^{\otimes
\frac{N}{2}}$. However, $E_{gl}$ can distinguish between GHZ and W
state since their distinctly different type of tangles leads to
different values of $\langle\tau\rangle$.

We consider the two seminal multi-qubit entangled state:
$|GHZ\rangle=\frac{1}{\sqrt{2}}(|00...0\rangle+|11...1\rangle)$ and
$|W\rangle=\frac{1}{\sqrt{N}}(|00...01\rangle+|00...10\rangle+...+|10...00\rangle)$
as initial states which are known to bear qualitatively different
quantum correlation. For example, GHZ state has a different tangle
distribution than that of W state. In operational context they
cannot be transformed to each other by local operation and classical
communication (LOCC) [20]. So it is interesting to investigate the
entanglement behavior of this two type of multi-qubit entangled
states under different models of system-environment interaction.

In the GHZ state, all the entanglement is contained in the N-tangle
form of non-local information which can be computed by N-concurrence
measure ($C_{N}$) [21]. $C_{N}$ is a generalization of two-qubit
measure, concurrence [11], for even number of qubits (N). Therefore,
for GHZ state GE is $E_{gl}=\tau_{N}=C_{N}^{2}$ . This can be easily
computed for even N as:
\begin{eqnarray}
R_{N}=\rho_{N}\sigma_{y}^{\otimes
N}\rho_{N}^{\ast}\sigma_{y}^{\otimes N}, \\
C_{N}=max\{\lambda_{1}-\sum_{i=2}^{2^{N}-1}\lambda_{i},0\};
\nonumber
 \end{eqnarray}
where $\sigma_{y}$ is Pauli matrix in y direction and
$\lambda_{i}^{,}$s are the eigenvalues of the matrix $R_{N}$ with
$\lambda_{1}$ being the maximum.

In the W state, only two-qubit entanglement, 2-tangles
($\tau_{2}$), is present with each qubit equally entangled  with
all other qubits. $\tau_{2}$ can be computed analytically from
the above measure. Thus, there are $\frac{N(N-1)}{2}$
different two-qubit entanglement in the W state. GE is therefore given by:
\begin{equation}
E_{gl}=(N-1)\tau_{2},
\end{equation}
for W state. It is important to note that this approach allows us
to obtain the exact solution of GE dynamics under decoherence of
different environment models. This also shows the great utility of
GE as a measure of entanglement.

\section{Decoherence Model Systems}

In order to evolve our chosen states under influence of decoherence
we use the Lindblad form of master equation [22],
\begin{equation}
\frac{d\rho}{dt}=\sum_{i=1}^{N}L_{i}\rho.
\end{equation}
The Lindblad operators, $L_{i}$, describe the local interaction of
each qubit with environment independent of other qubit interaction
with the environment. We assume $L_{i}$ is the same form for all
qubit, $L_{i}=L$. For markovian process [22]
\begin{equation}
L_{i}\rho=\sum_{k}\frac{\gamma_{k}}{2}[2J_{k}\rho
J_{k}^{\dag}-\{J_{k}J_{k}^{\dag},\rho\}],
\end{equation}
where operator $J_{k}$ describes the system-environment model of
interaction with strength $\gamma_{k}$. In this paper we investigate
dissipative, dephasing, and noise processes, each with a
well-defined $J_{k}$. For the two-level systems the operators,
${J_{k}}$, are expressed in terms of Pauli matrices. The solution of
Lindblad form of master equation for two-level systems are studied
in [23].

For dissipative environment, $J_{1}=\sigma_{-}$. In this process
the system interacts with a thermal bath at zero temperature. This
process could be described as spontaneous emission of a two-state
atom coupled with the vacuum modes of the ambient electromagnetic
field which leads the atom state to the ground state. For
dephasing process, $J_{1}=\sigma_{+}\sigma_{-}$. This is a
phase-destroying process that does not have a classical
counterpart and is therefore intrinsically quantum mechanical. It
corresponds to a situation where no energy is exchanged with
environment, that is, the  population of energy eigenstates of the
system do not change with time. Only the phase information which
includes quantum correlations is lost. For the noisy environment,
$J_{1}=\sigma_{-}$ and $J_{2}=\sigma_{+}$. Noisy dynamics are
related to another extreme of thermal bath, i.e. when temperature
is extremely high while the system-bath coupling is extremely
weak. This process randomizes the state of the system which
results in a completely mixed state eventually. The noise process
has a particular interest since its effect is basis independent.
That is, the noisy operation is invariant under unitary operation.
All these processes could have different effect on the multi-qubit
entangled state. But the common feature of them is that under the
action of each of these environments any initial entangled state
asymptotically evolves to a separable state.

\section{Results}

Our goal is to obtain the time dependence of the density matrix,
$\rho(t)=e^{-Lt}\rho(0)$, of the system in order to determine the
time evolution of $E_{gl}(t)$ for the initially prepared
multi-qubit entangled states, i.e. W and GHZ. According to the
structure of entanglement in W state we can deduce the time
dependency of the GE from two-qubit entanglement, $\tau_{2}$. The
two-qubit density matrix,
$\rho_{ij}(0)=\frac{2}{N}|\psi^{+}\rangle\langle\psi^{+}_{}|+\frac{(N-2)}{N}|00\rangle\langle00|
$, is the same for any pair of qubits, $ij$. Therefore, for the initial W state
in the  dissipative process we have,
\begin{equation}
\rho_{ij}(t)=\left(\begin{array}{cccc} \frac{N-2p}{N} & 0 & 0 & 0
\\0 & \frac{p}{N} & \frac{p}{N} & 0 \\ 0 & \frac{p}{N} &
\frac{p}{N} & 0
\\ 0 & 0 & 0 & 0
\end{array}\right),
\end{equation}
where $p=e^{-\gamma t}$ is the decoherence parameter. Similarly, for dephasing
process, one obtains,
\begin{equation}
\rho_{ij}(t)=\left(\begin{array}{cccc} \frac{N-2}{N} & 0 & 0 & 0
\\0 & \frac{1}{N} & \frac{p}{N} & 0 \\ 0 & \frac{p}{N} &
\frac{1}{N} & 0
\\ 0 & 0 & 0 & 0
\end{array}\right),
\end{equation}
and, therefore, for the initial W state, GE has the simple exact
solution,
\begin{equation}
E_{gl}(t)=\frac{4(N-1)}{N^{2}}e^{-2\gamma t},
\end{equation}
for both dissipative as well as dephasing processes. For noisy
process, the density matrix is
\begin{eqnarray}
\rho_{ij}(t)=\frac{2}{N}\left(\begin{array}{cccc}
\frac{1-p^{2}}{4} & 0 & 0 & 0
\\0 & \frac{1+p^{2}}{4} & \frac{p^{2}}{2} & 0 \\ 0 & \frac{p^{2}}{2}&
\frac{1+p^{2}}{4} & 0
\\ 0 & 0 & 0 & \frac{1-p^{2}}{4}
\end{array}\right)+ \\
\nonumber
\frac{(N-2)}{N} \left(\begin{array}{cccc} \frac{(1+p)^{2}}{4} & 0 &
0 & 0
\\0 & \frac{1-p^{2}}{4} & 0 & 0 \\ 0 & 0 &
\frac{1-p^{2}}{4} & 0
\\ 0 & 0 & 0 & \frac{(1-p)^{2}}{4}
\end{array}\right),
\end{eqnarray}
which leads to
\begin{equation}
E_{gl}=\frac{N-1}{4N^{2}}[max\{4p^{2}-(1-p^{2})^{\frac{1}{2}}(N^{2}-(pN-4p)^{2})^{\frac{1}{2}},0\}]^{2}.
\end{equation}
The dynamics of GE in dephasing, dissipation and noisy environment
for the initial W state is illustrated in Figs.1 and 2.
Eq.(10)(Fig.1) shows that the decay rate($\alpha$, for $E_{gl}
\propto \exp(-\alpha t)$ for GE is independent of N for the W state
in dissipative and dephasing environment as found previously using
numerical solution for a different measure of entanglement [10].
Note, however, that the rate of change of GE decreases with
increasing N. For noisy environment, Fig.2, we observe a decay to
separable state after a finite time $t_{sep}$ which increases
linearly with N, also consistent with previous studies [10].
\begin{figure}[h]
\includegraphics[width=9.0 cm]{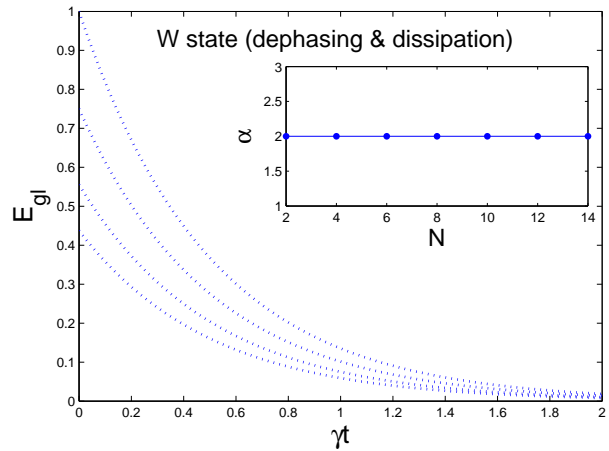}
\caption{$E_{gl}$ vs. $\gamma t$ in an initial W state with dephasing (or dissipative)
process for $N=2, 4, 6, 8$ qubits. The number of qubits increases
from the top curve to bottom. The inset shows the decay rate vs. $N$. Note
that these are simple graphs of Eq.(10) and are only drawn for comparison with other Figs.
}\label{1}
\end{figure}
\begin{figure}[h]
\includegraphics[width=9.0 cm]{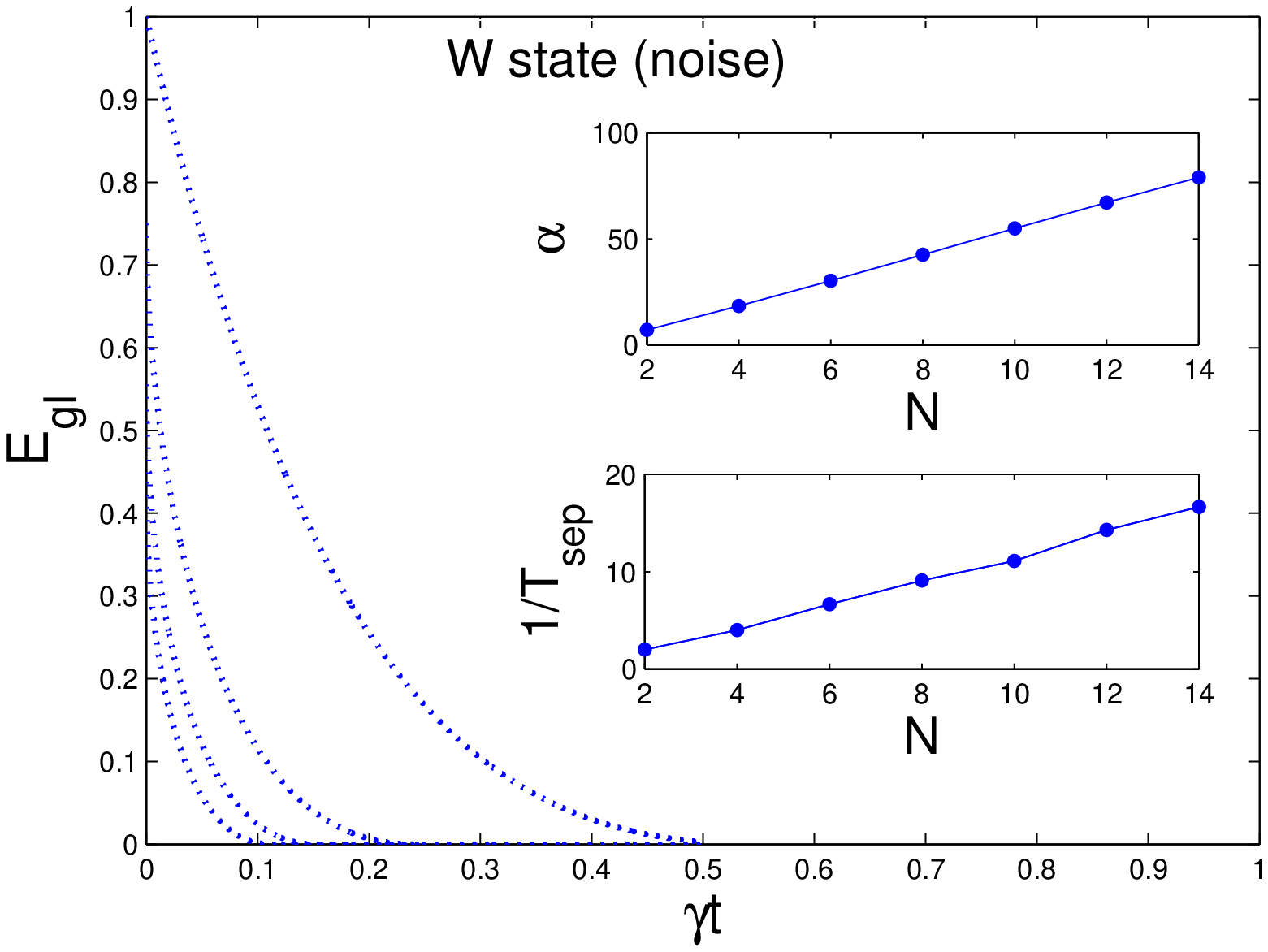}
\caption{$E_{gl}$ vs. $\gamma t$ in an initial W state with noisy
process for $N=2, 4, 6, 8$ qubits. The number of qubits increases
from the top curve to bottom. The inset shows the (linear) dependence of
decay rate on $N$ for up to $N=14$ qubits.
}\label{2}
\end{figure}

For the GHZ state, the density matrix in the dephasing process is
\begin{eqnarray}
\rho_{N}(t)=\frac{1}{2}(|0\rangle^{\otimes N}\langle0|^{\otimes N}+p^{\frac{N}{2}}|0\rangle^{\otimes N}\langle1|^{\otimes N}+\\
p^{\frac{N}{2}}|1\rangle^{\otimes N}\langle0|^{\otimes N}+
|1\rangle^{\otimes N}\langle1|^{\otimes N}),\nonumber
\end{eqnarray}
which leads simply to $E_{gl}=e^{-N\gamma t}$. The GHZ density
matrix for the dissipative process is
\begin{eqnarray}
\rho_{N}(t)=\frac{1}{2}(p^{\frac{N}{2}}|0\rangle^{\otimes N}\langle1|^{\otimes N}+p^{\frac{N}{2}}|1\rangle^{\otimes N}\langle0|^{\otimes N}\\
+ \sum_{q_{1},...,q_{N}=0}^{1}\lambda_{\langle
Z\rangle}|q_{1}q_{2}...q_{N}\rangle\langle
q_{1}q_{2}...q_{N}|),\nonumber
\end{eqnarray}
where
\begin{equation}
Z=\sum_{i=1}^{N}z_{i}\              \  ;\                  \
z_{i}= \frac{1-\sigma_{z_{i}}}{2}\nonumber,
\end{equation}
and
\begin{equation}
\langle Z\rangle=\langle
q_{1}q_{2}...q_{N}|Z|q_{1}q_{2}...q_{N}\rangle\nonumber,
\end{equation}
and
\begin{equation}
\lambda_{\langle Z\rangle}=[p^{\langle Z\rangle}(1-p)^{N-\langle
Z\rangle}+0^{\langle Z\rangle}].\nonumber
\end{equation}

The GHZ density matrix for the noisy process is
\begin{eqnarray}
\rho_{N}(t)=\frac{1}{2}(p^{N}|0\rangle^{\otimes N}\langle1|^{\otimes N}+p^{N}|1\rangle^{\otimes N}\langle0|^{\otimes N}\\
+ \sum_{q_{1},...,q_{N}=0}^{1}\lambda_{\langle
Z\rangle}|q_{1}q_{2}...q_{N}\rangle\langle
q_{1}q_{2}...q_{N}|),\nonumber
\end{eqnarray}
\begin{equation}
\lambda_{\langle Z\rangle}=\frac{1}{2^{N}}[(1+p)^{\langle
Z\rangle}(1-p)^{N-\langle Z\rangle}+(1-p)^{\langle
Z\rangle}(1+p)^{N-\langle Z\rangle}].\nonumber
\end{equation}
Consequently, our results ($E_{gl}$), for the GHZ state in
dephasing, dissipative and noisy environment are shown respectively
in Figs. 3, 4, and 5 for various N. As shown in the corresponding
insets, the decay rates increases linearly with system size N for
all three processes. Also, the rate of change of $E_{gl}$ increases
with system size as well. All these results are consistent with
previous studies using different methods than ours[9,10,11]. For the
dynamics of GE in GHZ state, although we have the exact solution
only for even number of qubit, the behavior of GE under decoherence
for the odd number can be inferred from the simplicity and symmetry
of our results. For example, our results for $E_{gl}$ in dephasing
process holds for any number of qubit in the initial GHZ state.
\begin{figure}[h]
\includegraphics[width=9.0 cm]{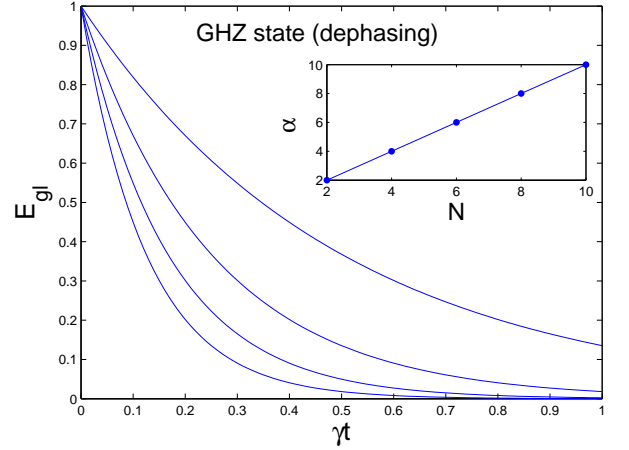}
\caption{$E_{gl}$ vs. $\gamma t$ in an initial GHZ state with dephasing
process for $N=2, 4, 6, 8$ qubits. The number of qubits increases
from the top curve to bottom. The inset shows the (linear) dependence of
decay rate vs. $N$ up to $N=10$ qubits.
}\label{3}
\end{figure}
\begin{figure}[h]
\includegraphics[width=9.0 cm]{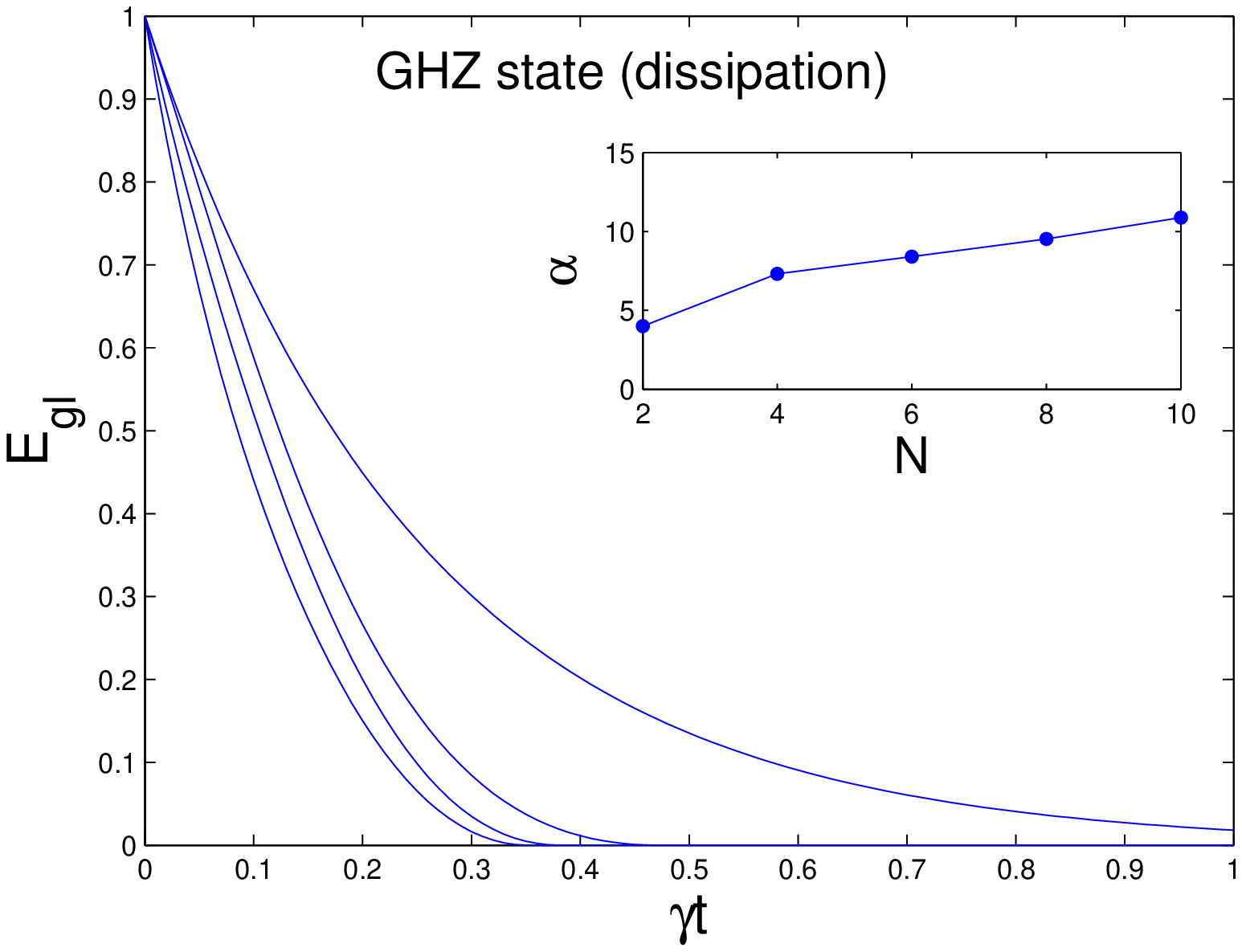}
\caption{$E_{gl}$ vs. $\gamma t$ in an initial GHZ state with
dissipation process for $N=2, 4, 6, 8$ qubits. The number of
qubits increases from the top curve to bottom. The inset shows
the (linear) dependence of decay rate on $N$ up to $N=10$ qubits.} \label{4}
\end{figure}
\begin{figure}[h]
\includegraphics[width=9.0 cm]{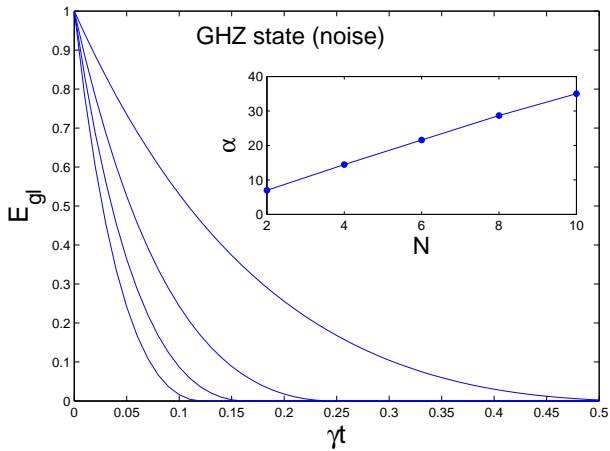}
\caption{ $E_{gl}$ vs. $\gamma t$ in an initial GHZ state with noisy
process for $N=2, 4, 6, 8$ qubits. The number of qubits increases
from the top curve to bottom. The inset shows the (linear) dependence
of decay rate vs. $N$ up to $N=10$ qubits.
}\label{5}
\end{figure}

\section{Conclusion}

In conclusion, in this work we have given an intuitive informational meaning
of MW measure of global entanglement.  Based on the relationship
between global entanglement and tangles, which constitute the
non-local form of the information, we identify the exact solution
of its dynamics under different system-environment
models for the two qualitatively different multi-qubit entangled
states, the GHZ and W states. In all the cases considered, we
obtain an exponential decay of entanglement as a function of time.
For the W state, the results show that the
lifetime of the GE is independent of the number of the qubits in
dephasing  and dissipative processes and the lifetime linearly decreases with N in
a noisy process. While for the GHZ state, the lifetime of GE
decreases linearly with N. Our results indicate that the quantum correlations
in W state are more robust to decoherence effects than that of the GHZ state.

The authors kindly acknowledge the support of Shiraz University Research Council.

%

\end{document}